# Extra-dimensional "metamaterials": simple models of inflation and metric signature transitions

Igor I. Smolyaninov

*Department of Electrical and Computer Engineering, University of Maryland, College Park, MD 20742, USA*

**Lattices of topological defects, such as Abrikosov lattices and domain wall lattices often arise as metastable ground states in higher-dimensional field theoretical models. We demonstrate that such lattice states may be described as extra-dimensional "metamaterials" via higher-dimensional effective medium theory. A 4+1 dimensional extension of Maxwell electrodynamics with a compactified time-like dimension has been considered as an example. It is demonstrated that from the point of view of macroscopic electrodynamics an Abrikosov lattice state in such a 4+1 dimensional spacetime may be described as a uniaxial hyperbolic metamaterial. Extraordinary photons perceive this medium as a 3+1 dimensional Minkowski spacetime in which one of the original spatial dimensions (the optical axis of the metamaterial) plays the role of a new time-like coordinate. Since metric signature of this effective space-time depends on the Abrikosov lattice periodicity, the described model may be useful in studying metric signature transitions. A particular kind of metric signature transition understood as a macroscopic medium effect may emulate cosmological inflation.**

Metric signature change events (in which a phase transition occurs between say (-,-,+,+) and (-,+,+,+) space-time signature) are being studied in many modified general relativity and quantum gravity theories (see for example ref.[1,2], and the references therein). Studying such events represents an obvious challenge for field theory. In general, it is predicted that a quantum field theory residing on a spacetime undergoing a signature change reacts violently to the imposition of the signature change. Both the total number and the total energy of the particles generated in a signature change event are formally infinite [1,2]. Therefore, such a metric signature transition shares some similarities with the cosmological big bang. Both events lead to creation of a Minkowski spacetime, and a large number of particles which populate this spacetime. Very recently we have described a hyperbolic metamaterial system, which undergoes a lower dimensional version of such a metric signature transition [3]. Hyperbolic metamaterials are artificial uniaxial materials in which the dielectric permittivity has different signs along different orthogonal directions. These metamaterials may be used to model a 2+1 dimensional Minkowski spacetime in which the role of time is played by one of the spatial coordinates. When a metamaterial is built and illuminated with a coherent extraordinary laser beam, the stationary pattern of light propagation inside the metamaterial may be treated as a collection of particle world lines, which represents a complete "history" of this 2+1 dimensional Minkowski spacetime [4]. It appears that in a very strong magnetic field physical vacuum itself behaves as a hyperbolic metamaterial [5-7].

Motivated by these recent developments, we propose a simple model of 3+1 dimensional metric signature change event, which is based on Maxwell electrodynamics residing on a 4+1 dimensional spacetime with a compactified time-like dimension. While conceptually simple and easy to analyze, our model may be extended to higher-dimensional gauge theories similar to modern extensions of the Kaluza-Klein theory (see for example [8]). As a starting point, we notice that lattices of topological defects, such as Abrikosov lattices [9] and domain wall lattices [10] often arise as metastable ground states in 4D and higher-dimensional field theoretical models. For example, as demonstrated by Chernodub [11], physical vacuum in a strong magnetic field develops an Abrikosov lattice of superconducting $\rho$ meson condensates. We will demonstrate that such lattice states may be described as extra-dimensional "metamaterials" via higher-dimensional effective medium theory.

Higher-dimensional extensions of Maxwell electrodynamics have been described for example in ref.[12]. We will apply such an extension to 4+1 dimensional spacetime with a compactified time-like dimension:

$$ds^2 = -dx_0^2 + dx_1^2 + dx_2^2 + dx_3^2 + dx_4^2 = -\frac{T^2}{4\pi^2}d\phi^2 + dx_1^2 + dx_2^2 + dx_3^2 + dx_4^2 \quad (1)$$

Maxwell's equations in 4+1D in differential form are [12]:

$$\partial_i F^{ki} = J^k \quad (2)$$

where $F^{ik}$ is the electromagnetic field tensor, $J^k$ is the current, and the Latin letters range is i=0, …, 4. The four-component electric field is $E^\alpha = F^{0\alpha}$, while the six-component magnetic field is $B^{\alpha\beta} = F^{\alpha\beta}$, where the Greek letters range is $\alpha$=1, …, 4. As a result, Maxwell's equations for the field components may be written as

$$\partial_\alpha E_\alpha = \rho \qquad e_{\alpha\beta\gamma} \partial_\alpha B_{\beta\gamma} = 0 \quad (3)$$

$$\partial_\alpha E_\beta - \partial_\beta E_\alpha = -\partial_0 B_{\alpha\beta} \qquad -\partial_\alpha B_{\alpha\beta} = \partial_0 E_\beta + J_\beta$$

where $e_{\alpha\beta\gamma}$ is the Levi-Civita symbol. Our goal is to introduce an effective medium description of the lattice media, which would be similar to macroscopic electrodynamics. Let us assume the most simple case in which the effective medium may be considered to be non-magnetic, so we only need to introduce one additional field $D_\alpha$ as an electric field averaged over macroscopic distances (much larger than the lattice periodicity). Thus, in the absence of "external" charges and currents our "macroscopic" field equations take the form

$$\partial_\alpha D_\alpha = 0 \qquad e_{\alpha\beta\gamma}\partial_\alpha B_{\beta\gamma} = 0 \qquad (4)$$

$$\partial_\alpha E_\beta - \partial_\beta E_\alpha = -\partial_0 B_{\alpha\beta} \qquad -\partial_\alpha B_{\alpha\beta} = \partial_0 D_\beta$$

These macroscopic field equations must be supplemented by the permittivity tensor $\varepsilon_{\alpha\beta}$ of the lattice medium:

$$D_\alpha = \varepsilon_{\alpha\beta} E_\beta \qquad (5)$$

This permittivity tensor is expected to be frequency dependent, so we will consider solutions of the form $e^{-i\omega_n x_0}$, where $\omega_n = 2\pi n/T$ is some Kaluza-Klein frequency. Thus, $\varepsilon_{\alpha\beta}$ of the lattice medium will be a function of *n*. Due to the symmetries of the Abrikosov lattices and the domain wall lattices considered here, $\varepsilon_{\alpha\beta}$ tensors of these media are diagonal:

$$\varepsilon_{\alpha\beta} = \begin{pmatrix} \varepsilon_1 & 0 & 0 & 0 \\ 0 & \varepsilon_1 & 0 & 0 \\ 0 & 0 & \varepsilon_1 & 0 \\ 0 & 0 & 0 & \varepsilon_4 \end{pmatrix} \qquad (6)$$

with different values $\varepsilon_4$ in the direction of the "optical axis", and $\varepsilon_1$ in the orthogonal directions, respectively. Numerical values of $\varepsilon_1$ and $\varepsilon_4$ for the typical lattice media will be evaluated later in this paper. We will consider propagation of "extraordinary waves" in such a medium. For the extraordinary waves $E_4 \neq 0$, while for the "ordinary waves" $E_4 \equiv 0$. We will define the extraordinary wave function as $\psi = E_4$ so that the ordinary portion of the electromagnetic field does not contribute to $\psi$. Solving the "macroscopic Maxwell's equations" (4) for $\psi = E_4$ we obtain

$$\varepsilon_2 \frac{\partial E_4}{\partial x_0} = -\frac{\partial B_{14}}{\partial x_1} - \frac{\partial B_{24}}{\partial x_2} - \frac{\partial B_{34}}{\partial x_3} \tag{7}$$

where $B_{\alpha\beta}$ may be found from the third equation in (4). After simple transformations the wave equation for the extraordinary field may be written as

$$-\frac{1}{\varepsilon_1}\frac{\partial^2 E_4}{\partial x_4^2} - \frac{1}{\varepsilon_4}\left(\frac{\partial^2}{\partial x_1^2} + \frac{\partial^2}{\partial x_2^2} + \frac{\partial^2}{\partial x_3^2}\right)E_4 = \omega_n^2 E_4 \tag{8}$$

We note that if $\varepsilon_1 > 0$, while $\varepsilon_4 < 0$, this wave equation coincides with the Klein-Gordon equation in 3+1 dimensional Minkowski spacetime. However, instead of compactified $x_0$ dimension, the role of time in eq.(8) is played by the spatial dimension $x_4$.

Let us calculate the values of $\varepsilon_1$ and $\varepsilon_4$ as a function of lattice periodicity for the typical lattice media shown in Fig.1. These calculations can be done in simple analytical form for the most important case of low volume concentration $n$ of the topological defects in vacuum. Let us assume that response of an individual topological defect to an external electric field may be characterized by dielectric permittivity $\varepsilon_m$ (its numerical value will be evaluated later in this paper.). Following [13], let us evaluate an integral

$$\frac{1}{V}\int (D_\alpha - E_\alpha)dV \equiv \langle D_\alpha \rangle - \langle E_\alpha \rangle \tag{9}$$

in which the integrand differs from zero only inside the defects. Therefore, this integral must be proportional to the volume concentration $n$ of the defects. For the domain wall lattice shown in Fig.1(a) field components $E_1$, $E_2$, $E_3$, and $D_4$ must be continuous. These requirements lead to

$$\langle D_4 \rangle = \langle E_4 \rangle \left(1 + n\left(1 - \frac{1}{\varepsilon_m}\right)\right) \quad \text{and} \quad \langle D_1 \rangle = \langle E_1 \rangle (1 + n(\varepsilon_m - 1)) \tag{10}$$

Thus, in the leading order the dielectric permittivity tensor components of the domain wall lattice can be estimated as

$$\varepsilon_4 \approx 1 \quad \text{and} \quad \varepsilon_1 \approx 1 + n\varepsilon_m \tag{11}$$

On the other hand, similar consideration of the Abrikosov lattice shown in Fig.1(b) requires $E_4$ to be continuous leading to

$$\varepsilon_1 \approx 1 \quad \text{and} \quad \varepsilon_4 \approx 1 + n\varepsilon_m \tag{12}$$

Equations (11) and (12) demonstrate that even in the small $n$ limit $\varepsilon_1$ and $\varepsilon_4$ may have opposite signs if $\varepsilon_m$ is large and negative. Let us evaluate the possible values of $\varepsilon_m$.

Dissipative behavior, like in the Ohm law, is inconsistent with the Lorentz symmetry of physical vacuum. Therefore, from the point of view of the electric conductivity properties, a ground state of the vacuum can either be a superconductor or an insulator [14]. An example of anisotropic superconducting behavior of physical vacuum in a strong magnetic field due to spontaneous formation of the Abrikosov lattice of $\rho$ meson condensates has been described in [11]. Hyperbolic metamaterial properties of this state have been demonstrated in [5]. Somewhat similar magnetic-field-induced Abrikosov lattice ground state in a 4+1 dimensional asymptotically Anti-de Sitter space has been found recently by Bu *et al.* [15] using holographic approach. Extra dimensional "holographic" superconductors and inhomogeneous Abrikosov-like states in these superconductors were also considered in much detail in [16,17]. Within the scope of holographic model the frequency-dependent conductivity of the superconducting state can be obtained as (see eq. (4.4) in [16])

$$\sigma = \frac{2a_2}{i\omega L_{eff}^4 a_0} + \frac{i\omega}{2} - i\omega Log L_{eff} \tag{13}$$

where $L_{eff}$ is the effective AdS radius, and $a_0$ and $a_2$ are the model integration constants. Therefore, in the low frequency limit the dielectric permittivity $\varepsilon_m$ of the holographic superconductor may be obtained as

$$\varepsilon_m(\omega) \approx \frac{4\pi\sigma}{i\omega} \approx 2\pi - 4\pi Log L_{eff} - \frac{8\pi a_2}{\omega^2 L_{eff}^4 a_0} \approx \varepsilon_\infty - \frac{\omega_p^2}{\omega^2} \qquad (14)$$

which looks analogous to the usual Drude model. Thus, $\varepsilon_m$ may indeed be large and negative in the low frequency limit leading to opposite signs of $\varepsilon_1$ and $\varepsilon_4$ of the Abrikosov lattice. As a result, a 3+1 dimensional Minkowski spacetime does appear as an effective "macroscopic" metric of the Abrikosov lattice residing in a 4+1 dimensional spacetime with a compactified time-like dimension. Such a construct, if real, would represent ultimate manifestation of Mach's principle (see for example [18]) – spacetime signature of vacuum would be defined by distribution of topological defects (matter) in the universe.

Combining eqs.(12) and (14) produces the following result for the dielectric permittivity tensor components of the Abrikosov lattice in the low frequency limit:

$$\varepsilon_1 \approx 1 \quad \text{and} \quad \varepsilon_4 \approx 1 + n\left(\varepsilon_\infty - \frac{\omega_p^2}{\omega^2}\right) \approx 1 - n\frac{\omega_p^2}{\omega^2} \qquad (15)$$

where the effective plasma frequency $\omega_p$ can be obtained from eq.(14). As can be seen from eq.(8), an effective macroscopic "metric signature change" is observed at the critical value of the Abrikosov lattice density

$$n_c = \frac{\omega^2}{\omega_p^2} \qquad (16)$$

Typically, Abrikosov lattice density depends on the magnitude of external magnetic field [11,14,15]. Therefore, inhomogeneous field distribution may lead to a metric signature transition. However, compared to models described in [1,2] such a metric signature change does not present a challenge. Calculating the total number and the total energy of particles generated in a signature change event will not lead to divergencies [3]. All the results will be regularized by the finite periodicity of the Abrikosov lattice. These calculations may be performed similar to [3] as follows. The number of photons emitted during the metric signature change transition can be calculated via the dynamical Casimir effect [19,20]. The total energy $E$ of emitted photons (per one polarization state) from a metric signature changing volume $V$ depends on the photon dispersion laws $\omega(k)$ before and after the transition (compare to eq.(3) from [20]):

$$\frac{E}{V} = \int \frac{d^4\vec{k}}{(2\pi)^4} \left( \frac{1}{2}\hbar\omega_1(k) - \frac{1}{2}\hbar\omega_2(k) \right) \quad (17)$$

where $\omega_1(k)$ and $\omega_2(k)$ are the photon dispersion relations inside the effective medium before and after the effective "macroscopic metric signature" change. Equation (17) is valid in the "sudden change" approximation, in which the dispersion law is assumed to change instantaneously (the detailed discussion of the validity of this approximation can be found in ref. [20]). Therefore, the number of photons per frequency interval emitted during the transition can be written as

$$\frac{dN}{Vd\omega} = \frac{1}{2}\left( \frac{dn_1}{d\omega} - \frac{dn_2}{d\omega} \right) \quad (18)$$

where $dn_i/d\omega$ are the photonic densities of states before and after the transition. At $k \gg k_{max}$, where $k_{max}$ is the inverse vector of the Abrikosov lattice, the photonic densities of state coincide. Therefore, the integral in eq.(17) remains finite, even though quite large: similar to [3], $dN/d\omega \sim k_{max}^4$. In the continuous medium limit in which $k_{max} \to \infty$ the number of photons emitted during the transition would diverge in agreement with results of [1,2].

It is also interesting to note that a particular kind of metric signature transition understood as a macroscopic medium effect may emulate inflation. This fact has been noted in ref. [21] in regards to 3D hyperbolic metamaterials. The metric of 3+1 dimensional inflationary de Sitter spacetime may be written as

$$ds^2 = -dx_4^2 + e^{Hx_4}(dx_1^2 + dx_2^2 + dx_3^2) \tag{19}$$

where the Hubble constant $H \sim \Lambda^{1/2}$ ($\Lambda$ is the cosmological constant). The corresponding Klein-Gordon equation is

$$-\frac{\partial^2 \varphi}{\partial x_4^2} + \frac{1}{e^{Hx_4}}\left(\frac{\partial^2 \varphi}{\partial x_1^2} + \frac{\partial^2 \varphi}{\partial x_2^2} + \frac{\partial^2 \varphi}{\partial x_3^2}\right) - H\left(\frac{\partial \varphi}{\partial t}\right) = \frac{m^2 c^2}{\hbar^2}\varphi \tag{20}$$

Now let us extend our 4+1 dimensional macroscopic electrodynamics model by letting adiabatic variations of $\varepsilon_1$ and $\varepsilon_4$ as a function of $x_4$. Taking into account spatial derivatives of $\varepsilon_1$ and $\varepsilon_4$ in eq.(4) results in the following modified equation for $E_4$:

$$-\frac{\partial^2 E_4}{\varepsilon_1 \partial x_4^2} + \frac{1}{(-\varepsilon_4)}\left(\frac{\partial^2 E_4}{\partial x_1^2} + \frac{\partial^2 E_4}{\partial x_2^2} + \frac{\partial^2 E_4}{\partial x_3^2}\right) + \left(\frac{1}{\varepsilon_1^2}\left(\frac{\partial \varepsilon_1}{\partial x_4}\right) - \frac{2}{\varepsilon_1 \varepsilon_4}\left(\frac{\partial \varepsilon_4}{\partial x_4}\right)\right)\left(\frac{\partial E_4}{\partial x_4}\right) +$$

$$+ \frac{E_4}{\varepsilon_1 \varepsilon_4}\left(\frac{1}{\varepsilon_1}\left(\frac{\partial \varepsilon_1}{\partial x_4}\right)\left(\frac{\partial \varepsilon_4}{\partial x_4}\right) - \left(\frac{\partial^2 \varepsilon_4}{\partial x_4^2}\right)\right) = \omega_n^2 E_4$$

(21)

It is easy to see that $\varepsilon_4$=const>0 and $\varepsilon_1 \sim -e^{-Hx_4}$ (where $x_4$ is considered a time-like variable) differs from eq.(20) only by a scaling factor in the limit of large $Hx_4$:

$$-\frac{\partial^2 E_4}{\partial x_4^2} + \frac{1}{\varepsilon_4 e^{Hx_4}}\left(\frac{\partial^2 E_4}{\partial x_1^2} + \frac{\partial^2 E_4}{\partial x_2^2} + \frac{\partial^2 E_4}{\partial x_3^2}\right) - H\left(\frac{\partial E_4}{\partial x_4}\right) = -\frac{\omega_n^2}{e^{Hx_4}}E_4 \approx 0 \qquad (22)$$

In this limit eq. (22) describes propagation of massless particles in the inflationary de Sitter metric described by eq.(19). According to eq.(11), this situation may occur in the lattice of domain walls in which $\varepsilon_4=1$, while $\varepsilon_1$ may be small and negative near the metric signature transition. In such a case the spatial distribution of domain walls should be

$$n(x_4) = \frac{1 + e^{-Hx_4}}{(-\varepsilon_m)} \qquad (23)$$

near the transition.

In conclusion, we have demonstrated that lattices of topological defects, which often arise as metastable ground states in higher-dimensional field theoretical models may be described as extra-dimensional "metamaterials" via higher-dimensional effective medium theory. A 4+1 dimensional extension of Maxwell electrodynamics with a compactified time-like dimension has been considered as an example. It appears that from the point of view of macroscopic electrodynamics an Abrikosov lattice state in such a 4+1 dimensional spacetime may be described as a uniaxial "hyperbolic metamaterial". Extraordinary photons perceive this medium as a 3+1 dimensional Minkowski spacetime in which one of the original spatial dimensions (the optical axis of the metamaterial) plays the role of a new time-like coordinate. Since metric signature of this effective space-time depends on the Abrikosov lattice periodicity, the described model may be useful in studying metric signature transitions. We also note that a particular kind of metric signature transition understood as a macroscopic medium effect may emulate cosmological inflation.

**Figure Captions**

**Figure 1.** Lattices of topological defects in a 4+1 dimensional spacetime considered in our model: (a) Lattice of domain walls, and (b) Abrikosov lattice

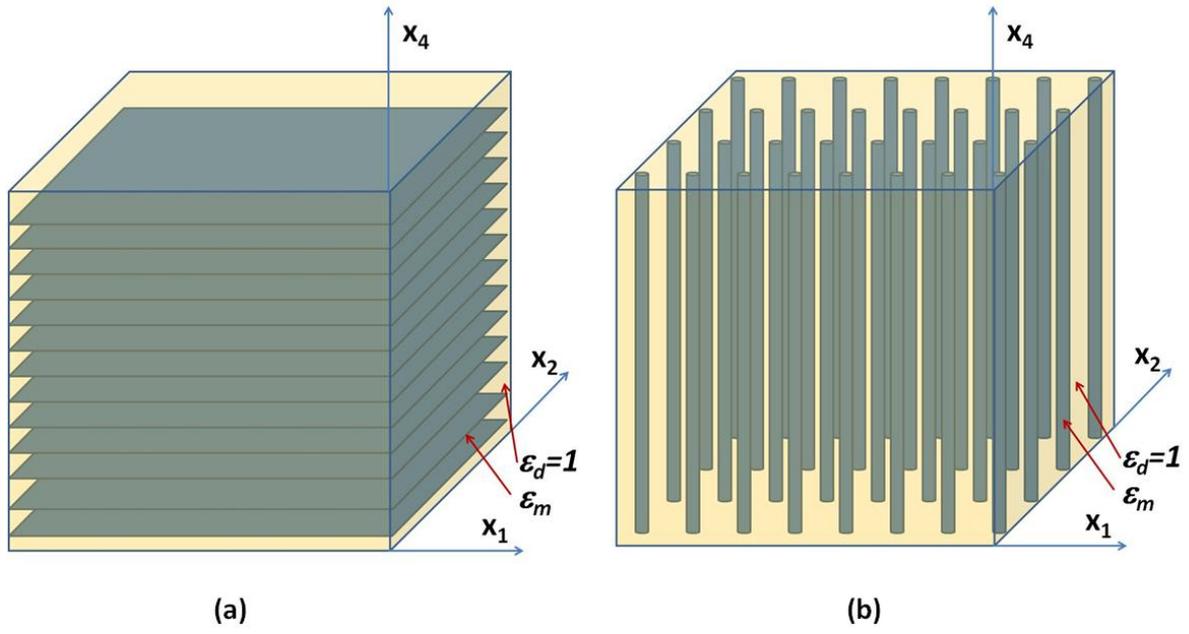

Fig.1